\def\leff{l_{\mbox{\scriptsize eff}}}
\def\ie{{\it i.e.}}

\def\O5{\Omega_5}

\def\Yo{{Y}}
\def\Ym2{{_{-2}Y}}
\def\Yp2{{_{2}Y}}
\def\Xolm{X_{1,lm}}
\def\Xtlm{X_{2,lm}}
\def\Xolmm{X_{1,l-m}}
\def\Xtlmm{X_{2,l-m}}
\def\Y{{\bf Y}}
\def\D{{\bf D}}
\def\P{{\bf P}}
\def\R{{\bf R}}
\def\RR{{\bf R}_{45}}
\def\lm{{lm}}
\def\elm{{E,lm}}
\def\blm{{B,lm}}
\def\elmm{{E,l-m}}
\def\blmm{{B,l-m}}
\def\tlm{{2,lm}}
\def\mtlm{{-2,lm}}
\def\lm{{lm}}
\def\al{\hbox {\boldmath $\alpha$}}
\def\N{{\bf N}}
\def\E{{\bf E}}
\def\B{{\bf B}}
\def\PP{{\bf\Pi}}
\def\e{{\bf e}}

\def\a{{\bf a}}
\def\b{{\bf b}}
\def\k{{\bf k}}
\def\q{{\bf q}}
\def\r{{\bf r}}
\def\th{\hbox{\boldmath $\theta$}} 
\def \I{\bf I}

\def \Nid{\sigma^2 \I}

\def\expec#1{\langle#1\rangle}
\def\eq#1{equation~(\ref{#1})}

\def\beq{\begin{equation}}
\def\eeq{\end{equation}}
\def\beqa{\begin{eqnarray}}
\def\eeqa{\end{eqnarray}}
\def\sec#1{Section~\ref{#1}}

\newcommand{\be}{\begin{equation}}
\newcommand{\ee}{\end{equation}}

\newcommand{\ra}{\;\raise1.0pt\hbox{$'$}\hskip-6pt\partial\;}
\newcommand{\lo}{\;\overline{\raise1.0pt\hbox{$'$}\hskip-6pt
\partial}\;}

\documentstyle[prd,aps,epsf]{revtex}

\begin{document}

\title{$E/B$ decomposition of finite pixelized CMB maps}
\author{Emory F. Bunn}
\address{Physics Department, University of Richmond, Richmond, VA  23173}
\author{Matias Zaldarriaga}
\address{Physics Department, New York University,
4 Washington Place, New York, NY 10003}
\address{Institute for Advanced Study, Einstein Drive,
Princeton, New Jersey, NJ 08540}
\author{Max Tegmark}
\address{Dept. of Physics, Univ. of Pennsylvania, Philadelphia, PA 19104}
\author{Angelica de Oliveira-Costa}
\address{Dept. of Physics, Univ. of Pennsylvania, Philadelphia, PA 19104}

\maketitle

\vskip 1pc


\begin{abstract}

Separation of the $E$ and $B$ components of a microwave background
polarization map or a weak lensing map is an essential 
step in extracting science from it, but when the map covers only 
part of the sky and/or is pixelized, this decomposition cannot be done perfectly.
We present a method for decomposing an arbitrary sky map into a sum of 
three orthogonal components that we term ``pure $E$'', ``pure $B$'' and 
``ambiguous''. The fluctuations in the pure $E$ and $B$ maps are due only to the 
the $E$ and $B$ power spectra, respectively, whereas the source of those in the  
ambiguous map is completely indeterminate.
This method is useful both for providing intuition for experimental design
and for analyzing data sets in practice.
We show how to find orthonormal bases for all three components
in terms of bilaplacian eigenfunctions, thus providing 
a type of polarized signal-to-noise eigenmodes that simultaneously separate both 
angular scale and polarization type.
The number of pure and ambiguous modes probing a characteristic 
angular scale $\theta$ scales as the map area over $\theta^2$ 
and as the map boundary length over $\theta$, respectively.
This implies that fairly round maps 
(with short perimeters for a given area) will yield the most efficient 
$E/B$-decomposition and also that the fraction of the information
lost to ambiguous modes grows towards larger angular scales.
For real-world data analysis, we present a simple matrix eigenvalue  
method for calculating nearly pure $E$ and $B$ modes in pixelized maps.
We find that the dominant source of leakage between $E$ and $B$ 
is aliasing of small-scale power caused by the pixelization, essentially
since derivatives are involved.
This problem can be eliminated by heavily oversampling the map, but is 
exacerbated by the fact that the $E$ power spectrum is expected to be much larger than the $B$ power spectrum and by the extremely blue power spectrum that CMB polarization
is expected to have. We found that a factor of 2 to 3 more pixels are needed in a polarization map to achieve the same level of contamination by aliased power than in a temperature map. Oversampling is therefore much more important 
for the polarized case than for the unpolarized case, which should be reflected in
experimental design.

\end{abstract}
\pacs{98.70.Vc,98.80.-k}

\section{Introduction}

Detecting polarization of the cosmic microwave background
(CMB) radiation has become one of the main goals of
the CMB community.  Numerous experimental groups are currently
searching for CMB polarization \cite{Keating01,staggs,hedman,peterson,angelica}.
CMB polarization can potentially offer a vast
amount of information about our Universe.  In general, polarization in
very sensitive to the ionization history of the Universe. For example,
on large scales it can provide insight into the way the Universe
reionized \cite{zalreio}. On degree scales, once the temperature
anisotropies are well measured, the predicted polarization can serve
as a test of how and when recombination happened and could potentially
lead to an important confirmation of the Big Bang model
\cite{peebles,landau}. Moreover, because the bulk of the polarization is
produced at the last-scattering surface, it should exhibit no
correlation on scales larger than about one degree unless there were
super-horizon perturbations at decoupling. Polarization can thus become
a good test of inflation \cite{sperzal}.

Most of the recent interest in polarization is based on its ability to
provide evidence for a stochastic background of gravity waves. It has
been shown that the polarization field on the sky can be decomposed
into two parts, a scalar part usually called $E$ and a pseudoscalar
part usually called $B$ \cite{2.kks,3.spinlong}. The pseudoscalar part
cannot be created by density perturbations to linear order in
perturbation theory. A detection of the $B$ component on large scales
would thus indicate the presence of a background of gravity waves, a
prediction of inflationary models \cite{spinlett,kkslett}. 
Such a detection would determine the energy scale of inflation and
could provide a stringent test of inflationary models
\cite{kinney}. On smaller scales, the $B$ modes will most probably be
dominated by secondary contributions produced after last scattering,
the leading one being gravitational lensing
\cite{pollens}. A detection of these contributions
could provide information about the
distribution of matter all the way up to the last-scattering
surface. There are many proposals for how to detect and use this
effect \cite{jacek,karim,wayne}.  In standard models, however,
the $B$ component is likely to be quite difficult to detect
\cite{JKW,maxangelica,ted}.

It is clear that a separation of the observed polarization into $E$
and $B$ parts is crucial to much of the CMB polarization scientific
program. It has been realized, however, that real-world complications
such as the finite size of the observed patch can significantly reduce
our ability to do a clean separation between the two components: when using a 
quadratic estimator method for measuring the $E$ and $B$ power spectra,
substantial ``leakage'' between the two was found on large angular scales
\cite{maxangelica}. In
 \cite{ted} it was shown that naive estimates of the sensitivity needed
to detect the $B$ component that ignore the such leakage can
significantly underestimate the required sensitivity for an experiment
aimed at detecting the $B$ modes. In \cite{lewis} it was shown that in
a finite patch, modes that are only $E$ or only $B$ can be constructed
but that there are also ambiguous modes, modes that receive
contributions to their power from both $E$ and $B$. The construction
of the modes was done for a round patch working in harmonic space. It
was shown for each value of $m$ there are two ambiguous modes.

The issue of separating $E$ and $B$ has also generated interest in
 the field of weak gravitational lensing\cite{kaiserlens,huwhitelens,Crittenden02},
where the basic cosmological signal is expected to produce only an $E$-pattern in
cosmic shear maps, and the $B$-mode therefore serves as an important test for other signals due 
to intrinsic galaxy alignment or systematic errors. Although we do not discuss weak lensing explicitly
in this paper, our results are relevant to that case as well since the lensing $E/B$ problem is 
mathematically analogous.

In this paper we revisit the issue of $E$ and $B$ mode separation, with two
goals: to provide intuition for experimental design
and for efficiently analyzing data sets in practice.
We present a general derivation of the pure $E$, pure $B$ and ambiguous modes in
real space, and relate them to the eigenfunctions of the bilaplacian on
a finite patch. We then introduce a way to obtain modes that are
very nearly ``pure'' in a
pixelized map by solving a generalized eigenvalue problem and discuss how this can be
used to analyze real-world data sets.

The paper is organized as follows.  Section II establishes some notation
and reviews the mathematics underlying the $E/B$ decomposition of a polarization
field.  In Section III, we show how to decompose the
space of all polarization fields on a finite patch of sky into
pure $E$ modes, pure $B$ modes, and modes that are ambiguous
with respect to the $E/B$ decomposition.  Section IV presents examples
of this decomposition.
In Section V, we present a method for finding
(nearly) pure $E$ and $B$ modes numerically
for pixelized maps by solving a generalized eigenvalue problem.  Section VI
presents examples.  In Section VII we show that aliasing of small-scale
power is the dominant source of ``leakage'' between the $E$-modes and the
$B$-modes.  We summarize our conclusions in Section VIII.

\section{$E$ and $B$ modes: Notation and preliminaries}

In this section we will review the definition of $E$ and $B$ modes to
introduce all the relevant notation. We will also give alternative
definitions of these modes which will help clarify how this
decomposition works on finite patches of sky. In \ref{flatsky} we
discuss the small-angle approximation.  Further details on properties
of spin-two fields on the sphere and the $E/B$ decomposition
may be found in, {\it e.g.}, \cite{zal,huwhite} and references therein.

\subsection{Spin two notation}

This section is rather technical. Since all intuitive aspects of our results can be
understood in terms of the much simpler formulae that apply in the flat-sky 
approximation, some readers may wish to skip straight to \sec{flatsky} and revisit
this section as needed.

The (linear) polarization of the CMB is described in terms of the Stokes
parameters $Q$ and $U$. The definition of $Q$ and $U$ depends on the
coordinate system chosen. In this subsection we review definitions
that are valid
for the full sky, so we will use spherical coordinates to
define $Q$ and $U$. 

We will follow the notation of \cite{3.spinlong}.  The Stokes parameters can
be combined to form a spin 2 $(Q+iU)$ and a spin $-$2 $(Q-iU)$
combination. In the full sky these combinations can be decomposed using
spin-2 harmonics, 
\beqa Q+iU= \sum_\lm a_\tlm \ \Yp2_\lm \ \ \ &;&\ \
\ Q-iU= \sum_\lm a_\mtlm \ \Ym2_\lm
\label{decomp}
\eeqa

It is natural to introduce a scalar ($E$) and a pseudoscalar ($B$) 
field to describe polarization. The expansion coefficients of these
two fields in (ordinary spin-0) spherical harmonics are
\beqa
a_\elm= -(a_\tlm+a_\mtlm)/2 \ \ \ &;& \ \ \
a_\blm= i(a_\tlm-a_\mtlm)/2 .
\label{eb}
\eeqa
On the sphere, these two functions completely characterize the
polarization field \cite{3.spinlong}. They are important physically because
cosmological density perturbations cannot create $B$ type polarization
while gravitational waves can \cite{2.kks,3.spinlong}.  On small scales $B$ polarization can be generated by lensing \cite{pollens}, and furthermore $B$ may turn out to be a good monitor of foreground contamination, although at the moment nothing is known about how different foregrounds contribute to $E$ or $B$. 
In terms of $a_\elm$ and $a_\blm$ the Stokes parameters can be written as
\cite{zalpol}
\beqa
Q= - \sum_\lm (a_\elm \ \Xolm + i a_\blm \ \Xtlm) \ \ \ &;&\ \ \  
U= - \sum_\lm (a_\blm \ \Xolm - i a_\elm \ \Xtlm), 
\label{qu1}
\eeqa
where $\Xolm=(\Yp2_\lm+\Ym2_\lm)/2$ and
$\Xtlm=(\Yp2_\lm-\Ym2_\lm)/2$. These functions satisfy 
$\Xolm^*=-\Xolmm$ and $\Xtlm^*=-\Xtlmm$ which together with 
$a^*_\elm=a_\elmm$ and $a^*_\blm=a_\blmm$ make $Q$ and $U$ real
quantities.  

The spin-2 harmonics in equation (\ref{decomp}) can be related to the
usual spin-0 spherical harmonics by means of two first-order differential
operators, the spin-raising ($\ra$) and
spin-lowering ($\lo$) operators \cite{3.spinlong}, which are defined in 
spherical coordinates by
\beqa
\ra &=& -\sin^s\theta\,\left[{\partial\over\partial\theta}+
i\csc\theta\,{\partial\over\partial\phi}\right]\sin^{-s}\theta,\\
\lo&=&-\sin^{-s}\theta\left[{\partial\over\partial\theta}-i\csc\theta
{\partial\over\partial\phi}\right]\sin^s\theta,
\eeqa
where $s$ is the spin of the function to which the operator is being applied.
When applied to the spin-weighted spherical harmonics, these operators
yield the following identities:
\beqa
\ra _{s}\Yo_\lm &=& [(l-s)(l+s+1)]^{1/2}\ _{s+1}\Yo_\lm \nonumber \\
\lo _{s}\Yo_\lm &=& -[(l+s)(l-s+1)]^{1/2}\ _{s-1}\Yo_\lm.
\eeqa 
In particular, the spin-0 and spin-2 harmonics are related as follows:
\beqa
\Yp2_\lm &=& [(l-2)!/(l+2)!]^{1/2} \ra\ra \Yo_\lm, \nonumber \\
\Ym2_\lm &=& [(l-2)!/(l+2)!]^{1/2} \lo\lo \Yo_\lm.
\label{ypym}
\eeqa

Another useful consequence of these relations is
\beq
\lo\lo\ra\ra \Yo_\lm = \ra\ra\lo\lo \Yo_\lm 
= {(l+2)!\over (l-2)!} \Yo_\lm
= (l+2)(l+1)l(l-1)\Yo_\lm,   
\eeq
or equivalently that when acting on spin-zero variables,
\beq
\lo\lo\ra\ra = \ra\ra\lo\lo = \nabla^2(\nabla^2+2),
\label{nabla4}
\eeq
since $\nabla^2$ corresponds to $-l(l+1)$ in spherical-harmonic space.

Equations (\ref{decomp}), (\ref{eb}) and (\ref{ypym}) can be combined to
obtain:
\beqa
Q+iU= \ra\ra(\psi_E+i\psi_B) \ \ \ &;& \ \ \ 
Q-iU= \lo\lo(\psi_E-i\psi_B) \nonumber \\
\psi_E=-\sum_\lm [(l-2)!/(l+2)!]^{1/2} a_\elm \Yo_\lm \ \ \ &;& \ \ \ 
\psi_B=-\sum_\lm [(l-2)!/(l+2)!]^{1/2} a_\blm \Yo_\lm.
\label{psieb}
\eeqa
Thus $Q$ and $U$ can be written in terms of second derivatives of the
scalar and pseudoscalar ``potentials'' $\psi_E$ and $\psi_B$, which
are directly related to $E$ and $B$. Equation (\ref{psieb}) is 
analogous to the fact that a vector field can be written as a sum of a
gradient and a curl component. The difference for spin-2 fields is
that one can write them as {\it second derivatives} of the scalar and
pseudoscalar potentials. 

We pause to note that the reason why $E$ and $B$ are the focus of
attention instead of $\psi_E$ and $\psi_B$ is partly a matter of
convention. Perhaps more importantly, $E$ and $B$ have the same
power spectrum on small scales as the Stokes parameters, while the
derivatives in equation (\ref{psieb}) imply that the power spectra of the
Stokes parameters and those of $\psi_E$ and $\psi_B$ differ by a factor
$(l-2)!/(l+2)! \sim l^{-4}$.

To clarify the relation between all these quantities, we can
think of weak gravitational lensing ({\it e.g.}, \cite{kaiserlens,huwhitelens}).
The shear variables are the
analogues of the Stokes parameters, $E$ is the analogue of the
projected mass density, and $\psi_E$ is the analogue of the projected gravitational
potential.

We can use equations (\ref{psieb}) and (\ref{nabla4}) to show that,
\beqa 
\nabla^2(\nabla^2+2) \psi_E &=& [\lo\lo(Q+iU) + \ra\ra (Q-iU)]/2\nonumber \\ 
\nabla^2(\nabla^2+2) \psi_B &=& i [\lo\lo(Q+iU) - \ra\ra
(Q-iU)]/2.  
\eeqa 
These equations show that we can take linear combinations of second
derivatives of the Stokes parameters
and obtain variables that depend only on $E$
or on $B$.  (In the flat-sky approximation, the left-hand sides 
of these equations are simply $\nabla^2E$ and $\nabla^2B$ respectively.   
On the
sphere, the relation is not so simple, but it is still true that the
left-hand sides depend only on $E$ and $B$ respectively.)
We will
use this to project out the $E$ and $B$ contributions.

\subsection{Vector notation}

We can summarize the above results using a slightly different
notation that will help clarify the analogy with vector fields.  
We will use boldface to denote the polarization field 
written in the form of a vector, $\P=\left(\matrix{Q\cr U}
\right)$. 
We then define two second-order differential operators 
$\D_B$ and $\D_E$, 
\begin{eqnarray}
\D_E&=&{1\over 2}\left(\matrix{\ra\ra+\lo\lo \cr
-i (\ra\ra-\lo\lo)}\right)\\
\D_B&=&{1\over 2}\left(\matrix{i(\ra\ra-\lo\lo) \cr
\ra\ra+\lo\lo}\right).
\end{eqnarray}
Equation (\ref{psieb}) now becomes
\beqa
\P=\D_E \psi_E + \D_B \psi_B,
\eeqa
the analogue of the gradient/curl decomposition. 
Moreover, $\D_E$ and $\D_B$ satisfy two important properties,
\beqa
\D_E^\dag\cdot\D_B&=&\D_B^\dag\cdot\D_E=0,\label{propdedb1}\\
\D_E^\dag\cdot\D_E&=&\D_B^\dag\cdot\D_B = \nabla^2(\nabla^2+2).\label{propdedb2}
\eeqa
Equation (\ref{propdedb1}) is the spin-two analogue of the familiar fact that
$\nabla\times\nabla=0$.
Substituting \eq{psieb} into \eq{propdedb1} implies that
if a polarization field on the
sky has only $E$ as a source, it should satisfy $\D_B^\dag\cdot \P=0$ and if
it is only due to a $B$ component is should satisfy $\D_E^\dag\cdot \P=0$.

In this vector notation, equation (\ref{qu1}) can be written as
\beqa 
\P &=& - \sum_\lm
a_\elm \ \Y_\elm + a_\blm \ \Y_\blm, \nonumber \\
\Y_\elm &=&\left(\matrix{\Xolm  \cr - i  \Xtlm} \right) \ \ ; \ \
\Y_\blm =\left(\matrix{i \Xtlm \cr \Xolm}  \right). 
\label{qu2}
\eeqa 

\subsection{Small-angle approximation}\label{flatsky}

In this subsection, we present some formulas valid in the 
small-angle (flat-sky) approximation. When working in this limit,
it is more natural to measure the Stokes
parameters with respect to a Cartesian coordinate system $(x,y)$
instead of the usual polar coordinate axis. 
In the flat-sky approximation, the differential operators reduce to simply
\begin{eqnarray}
\ra &=& -(\partial_x+i\partial_y),\\
\lo &=& -(\partial_x-i\partial_y),\\
\D_E&=&\left(\matrix{\partial_x^2-\partial_y^2\cr 2\partial_x\partial_y
}\right),\label{DEdefEq}\\
\D_B&=&\left(\matrix{-2\partial_x\partial_y \cr \partial_x^2-\partial_y^2
}\right).\label{DBdefEq}
\end{eqnarray}
Using the above expressions it is trivial to demonstrate that
$\D_B^\dag\cdot \D_E= \D_E^\dag\cdot \D_B= 0$ and that 
$\D_E^\dag\cdot\D_E=\D_B^\dag\cdot\D_B = \nabla^4$. 
In the flat-sky approximation, $|\nabla^2|\gg 1$ (that is,
only modes with eigenvalues much greater than one contribute
significantly), so the $\nabla^2(\nabla^2+2)$ operator in 
equation (\ref{propdedb2}) has reduced to the bilaplacian $\nabla^4$. 

$\D_E$ and $\D_B$ are the spin-2 analogues of the familiar gradient and curl 
operators. Applying $\D_E$ or $\D_B$ to a scalar field gives $E$ and $B$ fields 
that have vanishing ``curl'' and ``gradient'', respectively.
Equations (\ref{DEdefEq}) and (\ref{DBdefEq}) show that 
$\D_B=\R\D_E$, where the 
$2\times 2$ matrix 
\beq\label{RdefEq}
\R\equiv\left(
\begin{array}{cc}
0 & -1 \\
1 & 0 \\
\end{array}
\right)
\eeq
simply performs a rotation taking $Q\mapsto -U$ and $U \mapsto Q$.
When drawing polarization fields as two-headed arrows
with length $(Q^2+U^2)^{1/2}$ and angle $\tan^{-1}(U/Q)/2$, 
this corresponds to rotating the polarization direction by $45^\circ$ at each point .
In other words, rotating the polarization directions of an $E$-field by $45^\circ$ 
gives a $B$-field.

The analogue of equation (\ref{qu2}) is now given in terms of Fourier modes,
\beqa
\P(\r) &=&  \int {d^2 k \over (2\pi)^2}
\left[E(\k ) \left(\matrix{\cos 2\phi  \cr \sin 2\phi}
\right) + B(\k ) \left(\matrix{-\sin 2\phi  \cr \cos 2\phi}
\right)\right]
e^{i\k\cdot\r},\quad
\r=\left({x\atop y}\right),\quad
\k=k\left({\cos\phi\atop\sin\phi}\right).
\label{fourier}
\eeqa   
In other words, the $E/B$ decomposition becomes local in Fourier space:
the  polarization direction of the $E$-component is parallel or perpendicular to 
$\k$ whereas that of the $B$-component makes a 45$^\circ$ angle with $\k$.

\section{A natural basis for polarization fields}

On a manifold without boundary, any polarization field
can be uniquely separated into an $E$ part and a $B$ part.  But
if there is a boundary ({\it i.e.}, if only
some subset $\Omega$ of the sky has been observed), 
this decomposition is not unique. Let us first introduce some notation to clarify the problem.

Polarization fields living on $\Omega$ form a normed vector space with the inner product 
\beq\label{InnerProductEq}
(\P,\P')\equiv \int_\Omega \P\cdot\P' d\Omega,
\eeq
and we say that two fields $\P$ and $\P'$ are orthogonal if $(\P,\P')=0.$
We refer to a polarization field $\P$ as 
\begin{itemize}
\item $E$ if it has vanishing curl, \ie,  $\D_B^\dag\cdot\P=0$,
\item $B$ if it has vanishing divergence, \ie,  $\D_E^\dag\cdot\P=0$,
\item pure $E$ if it is orthogonal to all $B$-fields, and
\item pure $B$ if it is orthogonal to all $E$-fields.
\end{itemize}
As long as $\Omega$ is simply connected, which we shall assume throughout
this paper, an equivalent definition of an $E$ polarization field
is one that can be derived from a potential $\psi_E$ via ${\bf P}={\bf D}_E
\psi_E$. (And, of course, an analogous statement holds for $B$ fields.
As always, the analogy with the more familiar case of vector fields
holds: any curl-free field is the gradient of a potential.)

On the complete sky, every polarization field can be uniquely represented
as a linear combination of an $E$ field and a $B$ field, and all $E$ fields
are perpendicular to all $B$ fields.  In other words, the space of all
polarization fields is the direct sum of two orthogonal subspaces: 
the space of all
$E$ fields and the space of all $B$ fields.  
[One way to prove these assertions is simply to use the
$E$ and $B$ spherical harmonics defined in (\ref{qu2}) as a basis.]
In this case, there is no distinction between an $E$ field
and a ``pure $E$'' field.

But
if only
some subset of the sky has been observed, so that $\Omega$ is a manifold
with boundary, then
this decomposition is not unique.
One way to see this is to note that there are modes that satisfy
both the $E$-mode and $B$-mode conditions simultaneously.
When we split a polarization field into an $E$ part and a $B$ part,
these ``ambiguous'' modes can go into either component.
In order to make the $E/B$ decomposition unique, we must first
project out the ambiguous modes.  

In other words, the subspaces of all $E$ modes and all $B$ modes
are no longer orthogonal: in fact, they overlap.
To recapture orthogonality, we must restrict our attention to the
pure $E$ and $B$ subspaces.  To be specific, the space of pure $E$ modes
is the orthogonal complement of the space of all $B$ modes, which includes
both pure $B$ modes and ambiguous modes.  Similarly, the space of pure
$B$ modes is orthogonal to both the pure $E$ modes and the ambiguous modes.
In summary, we can represent the space of all polarization fields
on $\Omega$ as a direct sum of three subspaces: pure $E$, pure $B$, and
ambiguous.

In this section, we show explicitly
how to construct orthonormal bases of pure $E$ modes, pure $B$ modes, and
ambiguous modes, so that we can
unambiguously decompose any polarization field into these three
components.  In \cite{lewis} this construction was presented for a cap
working in harmonic space. We here present the general formalism in
real space. For simplicity, we work in the flat-sky approximation,
although the construction works without this assumption.

We first construct the ambiguous modes.
An ambiguous mode $\P$ must be an $E$-mode, so $\P=\D_Ef$
for some scalar field $f$.  And it must also satisfy the $B$-mode condition:
$\D_E^\dag\cdot\P=0$.  Combining these, we get
\be
0=\D_E^\dag\cdot\D_E f = \nabla^4f.
\ee
So we can make a pair of ambiguous modes $\D_Ef$ and $\D_Bf$ 
out of any function $f$ that
satisfies $\nabla^4f=0$. 
All such biharmonic functions are determined by their values
and first derivatives on the boundary
of the region, so it is straightforward to form a basis of
them simply by choosing a basis for the set of scalar
functions on the boundary.

In the quest of separating the $E$ and $B$ contributions
the ambiguous modes are not very useful,
since we cannot know whether they are due to a cosmological $E$ or $B$ signal.
If we are willing to assume (on either observational
or theoretical grounds)
that $E$ dominates over $B$ on the angular scale of interest, then it may
be sensible to assume that power found in the ambiguous modes is
$E$ power.  This does enhance the accuracy with which the
$E$ power spectrum can be detected in a given data set \cite{ted}.

Of much more use are the ``pure'' $E$ and $B$ modes.
We now give an explicit construction
of these pure modes.

Let the scalar field $\psi_E$ generate
a pure $E$-mode $\D_E\psi_E$, and let $\D_B\psi_B$ 
be {\it any} $B$-mode
(not necessarily pure).  The requirement for a pure $E$-mode is
that these be orthogonal:
\be
\int_\Omega d^2r\, (\D_E\psi_E)\cdot(\D_B\psi_B)=0.
\ee
If we use the explicit forms (\ref{DEdefEq}) and (\ref{DBdefEq}) 
for the differential operators and integrate by parts twice to
move $\D_E$ over to the $\D_B\psi_B$ term,
this reduces to a line integral around the boundary of $\Omega$.
(After integrating by parts, the surface integral vanishes because it contains 
$\D_E^\dag\cdot\D_B\psi_B$, which is zero.)  The line integral contains
terms proportional to $\psi_E$ and $\hat{\bf n}\cdot\nabla\psi_E$.
The conditions for a pure $E$ mode are therefore
\begin{enumerate}
\item $\psi_E=0$ on the boundary $\partial\Omega$.
\item $\hat {\bf n}\cdot\nabla \psi_E=0$ on the boundary $\partial\Omega$.
\end{enumerate}

In other words, $\psi_E$ must satisfy both Dirichlet and Neumann boundary
conditions simultaneously.  Fortunately, the bilaplacian
operator has a complete set of eigenfunctions that satisfy
these boundary conditions.  To form an orthogonal
basis of pure $E$ modes, all we have to do is find a complete
set of such
eigenfunctions
and apply the operator
$\D_E$ to them.  Similarly, if we apply $\D_B$, we 
will have an orthogonal set of pure $B$-modes.
The pure $B$ modes can also be found by taking the pure $E$ modes
and rotating the polarization at each point by 45$^\circ$.

The boundary conditions turns out to have a simple geometrical
interpretation: for a pure $E$-mode the polarization on the
boundary must be parallel or perpendicular to the boundary;
for a pure $B$-mode it must make a 45$^\circ$ angle with the boundary.

The proof that these basis functions are orthogonal
is similar to the more familiar situation with
eigenfunctions of the Laplacian.  Let $f$ and $g$ be 
eigenfunctions of $\nabla^4$ with eigenvalues $\lambda$ and $\mu$,
and let them satisfy the boundary conditions.  Then $\D_Ef$
and $\D_Eg$ are two of our ``pure $E$'' basis functions.
Their inner product is
\be
\int_\Omega d^2r\,\D_Ef\cdot\D_Eg
=\int_\Omega d^2r\,f\nabla^4g=\mu\int_\Omega d^2r fg,
\ee
where we have integrated by parts twice and used the boundary condition
on $f$ to drop the boundary terms.
Of course the same argument with $f$ and $g$ switched
leads to the conclusion that the inner product is $\lambda$ times
the integral of $fg$.  If $\lambda\ne\mu$, then the integral must therefore vanish, and 
if $\lambda=\mu$, we can take a 
linear combination that orthogonalizes the two modes.
We choose to normalize all modes $\P$ so that $(\P,\P)$=1.

In conclusion, the pure $E$-modes, pure $B$-modes and ambiguous modes
form a complete orthonormal  basis for the space of all square-integrable
(\ie, $(\P,\P)<\infty$) polarization fields $\P$ in a sky region $\Omega$.
We found that a polarization field is 
\begin{itemize}
\item pure $E$ if it has vanishing curl and is parallel or perpendicular to the boundary,
\item pure $B$ if it has vanishing divergence and makes a $45^\circ$ angle with the boundary, and
\item ambiguous if it has vanishing divergence {\it and} curl.
\end{itemize}
These conclusions apply not only to the eigenmodes that we have
constructed but more generally, by linearity, to any field.  This
means that we can optionally decompose a polarization field $\P$ into
its three components directly, without going through the step of
expanding it in eigenmodes.  The pure $E$-component $\P_E$ is obtained
by solving the bi-Poisson equation $\nabla^4\psi_E=\D_E^\dag\cdot\P$
with Dirichlet and Neumann boundary conditions and computing
$\P_E=\D_E\psi_E$.  The pure $B$-component $\P_B$ is obtained
analogously, and the ambiguous component $\P_?$ is simply the
remainder, \ie, $\P_?=\P-\P_E-\P_B$.

\section{Worked examples I}

In this section we illustrate the above construction for
two worked examples:
a disk in the flat-sky approximation and a spherical cap.

\subsection{Disk}\label{disk}

Suppose that the observed region is a disk of radius $R$ with $R\ll 1$
radian, so that the flat-sky approximation is appropriate.

We begin
with the ambiguous modes.
We want to find functions $f$ with $\nabla^4f=0$.  
Assume a separable solution $f(r,\phi)=F(r)e^{im\phi}$.
We know that $\nabla^2(\nabla^2 f)=0$, so $\nabla^2 f$
must be a harmonic function.  The most general solution is
$\nabla^2f\propto r^me^{im\phi}$.
Solving this equation for $f$, we get two independent solutions:
\be
f(r,\phi)\propto\cases{r^me^{im\phi}\cr r^{m+2}e^{im\phi}}.
\label{famb}
\ee
As we expected, there are in general two solutions per $m$ (since there
are two conditions we wish to impose on the boundary).  Each solution
yields two ambiguous modes, ${\bf D}_Ef$ and ${\bf D}_Bf$, which turn
out to be just rotations of each other.  

In the case
$m=0$, though, these two solutions do not yield
any ambiguous modes, as ${\bf D}_Ef={\bf D}_Bf=0$.
The same is true for the first of the
two $m=1$ solutions, so there is only one pair of ambiguous modes with $m=1$.
This counting of modes agrees with \cite{lewis}.

We now proceed to find the pure $E$- and pure $B$-modes.  One way to construct
eigenfunctions of $\nabla^4$ is to take 
\be
f=a_\lambda + a_{-\lambda},
\ee
where $a_{\lambda}$ 
is an eigenfunction of $\nabla^2$ with eigenvalue $\lambda$ and
$a_{-\lambda}$ has eigenvalue $-\lambda$.  (These two are
obviously degenerate eigenvalues of $\nabla^4$ with eigenvalue
$\lambda^2$, so we can take linear combinations of them.  Of course
there are no well-behaved eigenfunctions of $\nabla^2$
with positive
eigenvalue over an entire manifold, but there are over
a manifold with boundary.)

\begin{figure}[p]
\centerline{\epsfbox{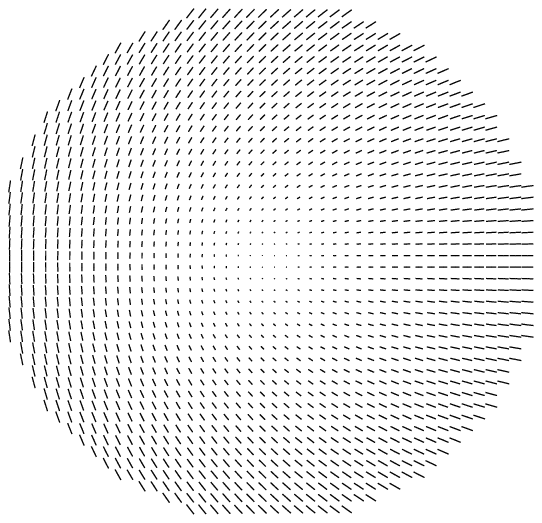}}
\centerline{\epsfbox{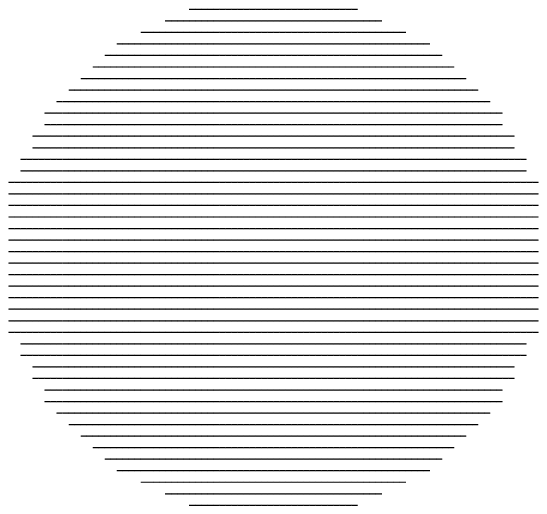}
\epsfbox{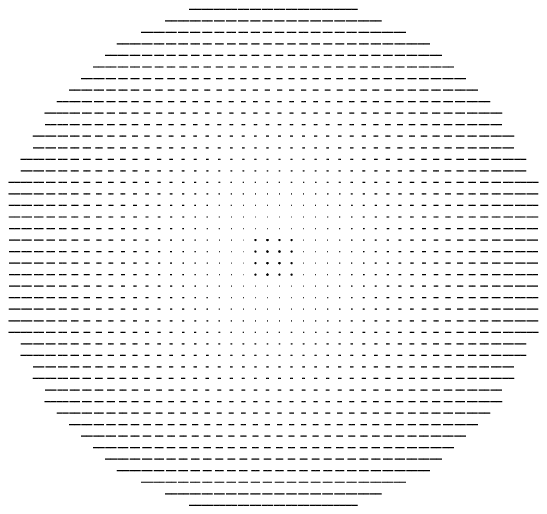}}
\centerline{\epsfbox{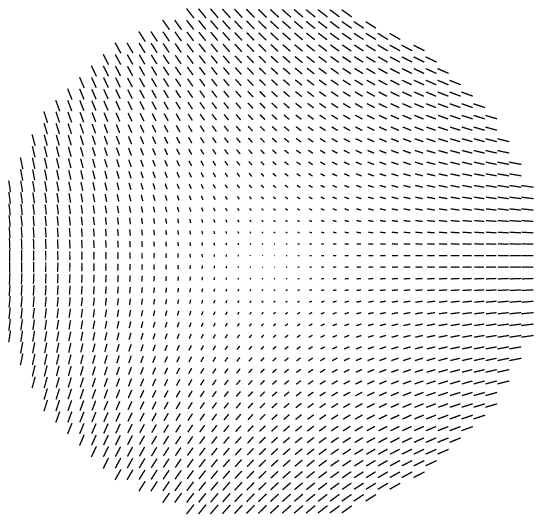}
\epsfbox{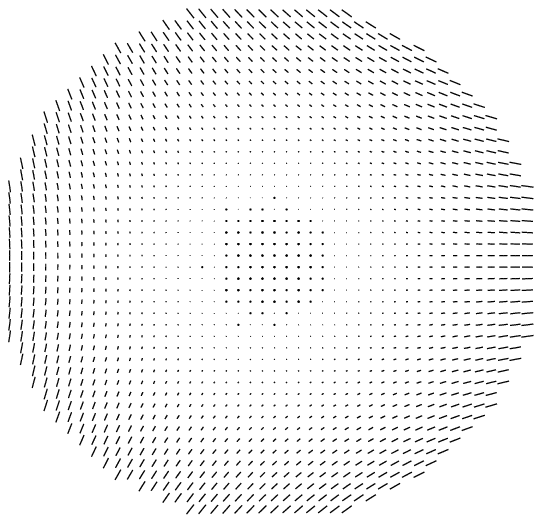}}
\caption{Ambiguous modes of a disk.  From top to bottom, $m=1,2,3$.}
\end{figure}

Once again we apply separation of variables in polar coordinates.
The angular dependence is $e^{im\phi}$.
Then for any
positive $k$, the Bessel function $J_m(kr)$ has eigenvalue $-k^2$
and the modified Bessel function $I_m(kr)$ has eigenvalue $k^2$, so
we can take our eigenfunctions of $\nabla^4$ to be
\be
f_{mk}(r,\phi) = (a J_m(kr)+ b I_m(kr))e^{im\phi}.
\label{eq:lincomb}
\ee
The boundary conditions tell us that
\be
{a\over b}=-{I_m(kR)\over J_m(kR)}=-{I_m'(kR)\over J_m'(kR)}.
\ee
So there will be solutions for all values of $k$ 
that satisfy $J_m'/J_m=I_m'/I_m$.  
These roots can be computed numerically.  For large
$n$, a good approximation for the $n$th root 
with azimuthal quantum number $m$ is
\be
k_{mn}R = \pi\left(n+{m\over 2}\right).
\ee

The figures show the first few modes of each type. As
noted above, there are no ambiguous modes with $m=0$, one
pair of ambiguous modes with $|m|=1$, and two for each $|m|>1$. Only
one of each pair is shown; the other is found by rotating the whole
pattern.  Similarly, for each pure $E$ and pure $B$ mode, a linearly
independent mode can be obtained by rotating the page.

\begin{figure}
\centerline{\epsfbox{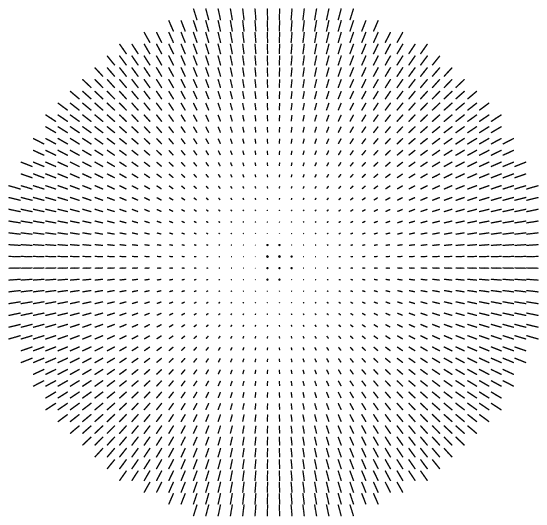}\epsfbox{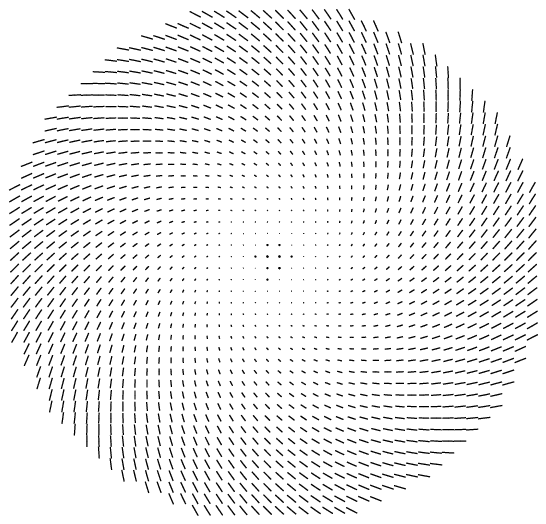}}
\centerline{\epsfbox{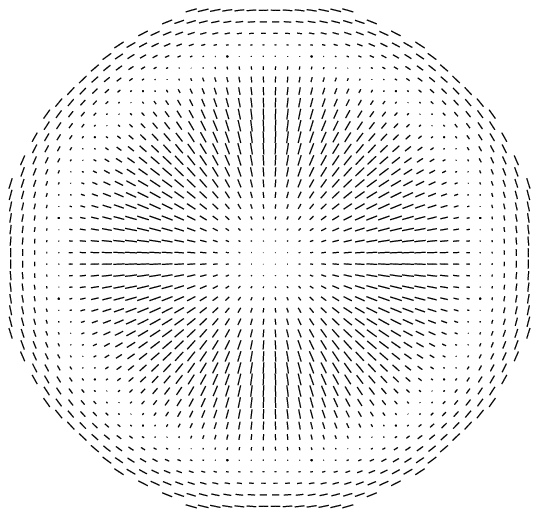}\epsfbox{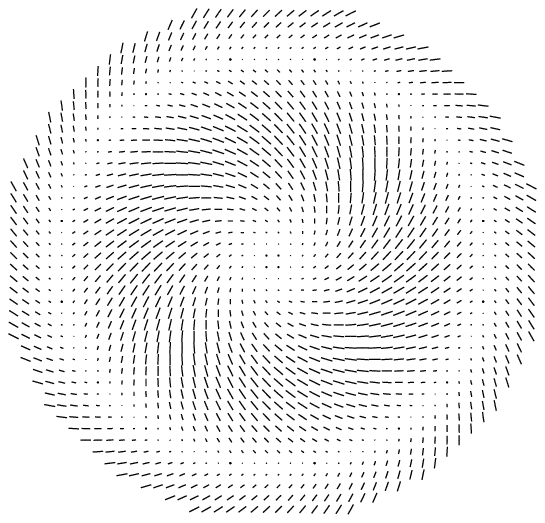}}
\caption{The first two $m=0$ pure $E$ (left) and pure $B$ (right) modes
for a disk.}
\label{fig:pure1}
\end{figure}
\begin{figure}
\centerline{\epsfbox{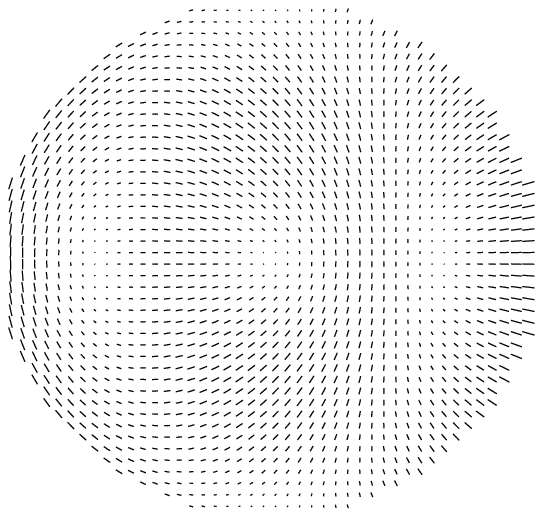}\epsfbox{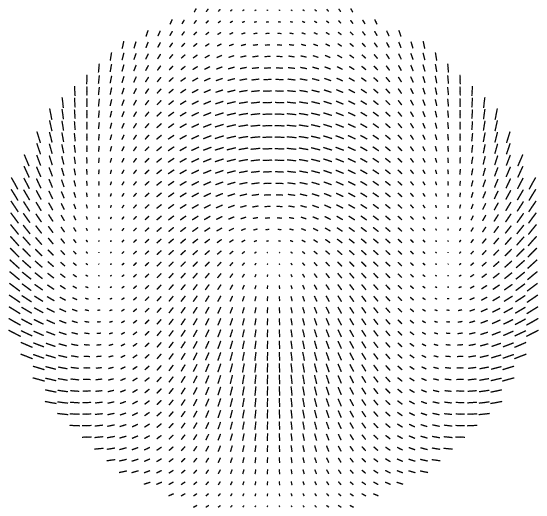}}
\centerline{\epsfbox{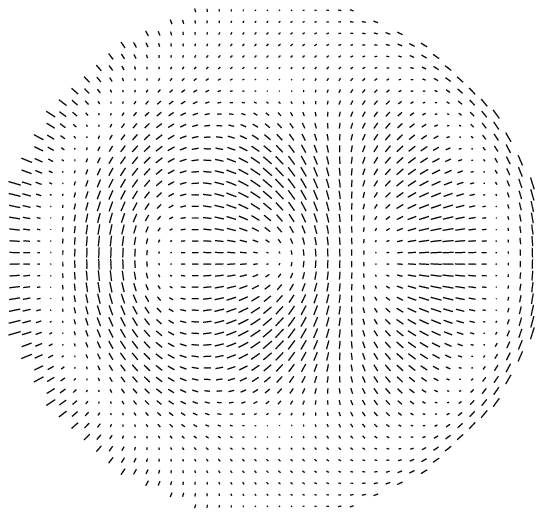}\epsfbox{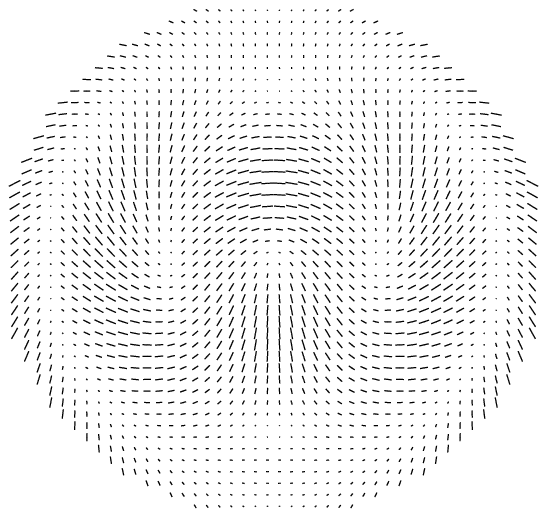}}
\caption{Same as Fig. \ref{fig:pure1} with $m=1$.}
\end{figure}
\begin{figure}
\centerline{\epsfbox{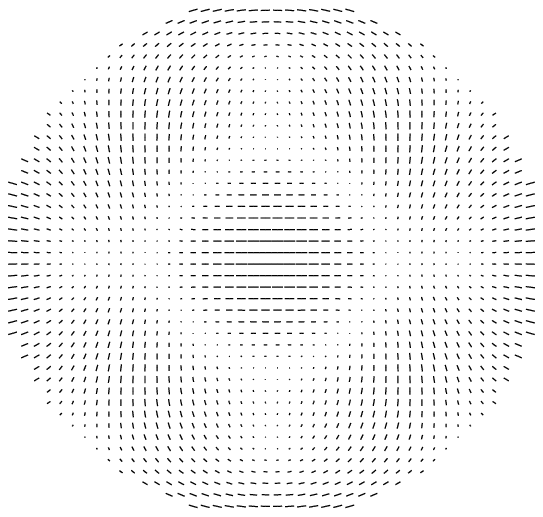}\epsfbox{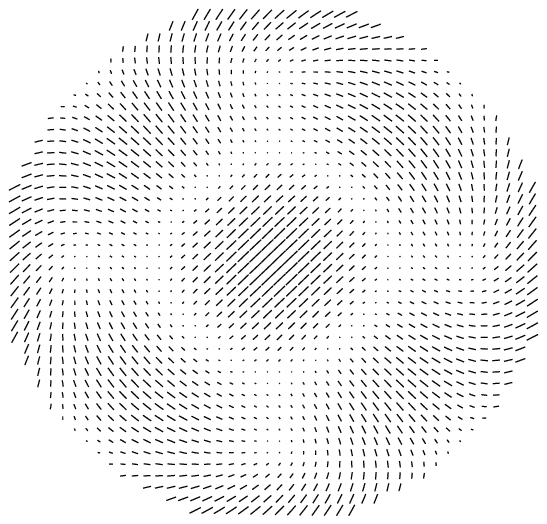}}
\centerline{\epsfbox{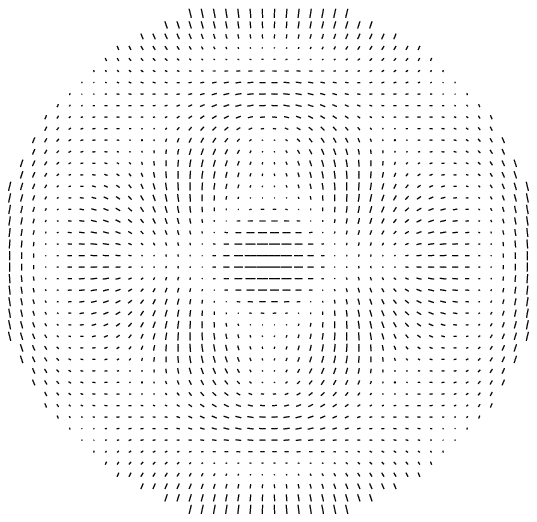}\epsfbox{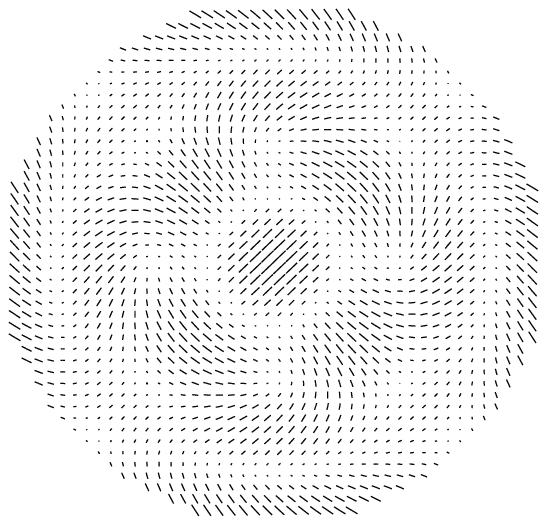}}
\caption{Same as Fig. \ref{fig:pure1} with $m=2$.}
\end{figure}

If our data covered the entire plane, we would construct a
basis out of only the ordinary Bessel functions $J_m$, excluding the
modified Bessel functions $I_m$.  In the limit $kR\to\infty$,
therefore, we expect the contribution from $I_m$ to be small, and
indeed this is the case.  The function $I_m$ grows exponentially for
large argument, so in order to satisfy the Dirichlet boundary
condition the coefficient $b$ in equation (\ref{eq:lincomb}) must be
small.  For a mode with $kR\gg 1$, therefore, the ordinary Bessel
function dominates except near the boundary.  In this limit, the
modified Bessel function takes over in a small region near the
boundary to ``flatten out'' the mode and make it satisfy the Neumann
boundary condition.

It is worth noting that all modes except those with $m=0$ require
that both $Q$ and $U$ be measured in the patch. Modes with $m=0$
depend only on $Q$ for the pure $E$ modes and only on $U$ for the pure
$B$ modes (with $Q$ and $U$ defined with respect to the polar
coordinates).

\subsection{Spherical cap}

This construction can be adapted to give the basis functions
for a spherical cap without recourse to the flat-sky approximation.
In this case, the functions we are looking for are eigenfunctions
of the operator 
\be 
\nabla^2(\nabla^2+2)=(\nabla^2+1)^2-1.
\label{eq:factorization}
\ee

The ambiguous modes will therefore be eigenfunctions
of the Laplacian with eigenvalues 0 and $-2$.  These eigenfunctions 
can be written in terms of associated Legendre functions $P_{lm}$ as
\be
f_{\rm amb}(\theta,\phi)=\cases{P_{0m}(\cos\theta)e^{im\phi}\cr
P_{1m}(\cos\theta)e^{im\phi}}
\ee
for any integer $m$.

The associated Legendre function $P_{lm}$ is well-behaved over
the entire sphere as long as $|m|\le l$, so there appear to be
four singularity-free solutions over the entire sphere.
These are mapped to zero
by $\D_E$ and $\D_B$, though, so they do not give 
ambiguous modes.  This is of course as it should be: there
are no ambiguous modes over the entire sphere.  

If, however, the region of interest $\Omega$ 
is a spherical cap $\theta\le\Theta$,
then we permit functions that have singularities outside $\Omega$.
In that case, there is one nontrivial ambiguous mode
with $m=\pm 1$, namely $P_{0\pm 1}e^{\pm i\phi}$, and two for every
$m$ with $|m|>1$.  The $l=0$ modes can be written explicitly as
\be
P_{0m}(\cos\theta)=
\left(\sin\theta \over 1+\cos\theta\right)^m.
\ee
The $l=1$ modes are not so simple.  The first one is
\be
P_{12}(\cos\theta)={(\cos\theta+2)\sin^2\theta\over(1+\cos\theta)^2},
\ee
and the remainder can be computed from recurrence relations.

We can also construct the pure $E$ and $B$ modes from the associated
Legendre functions.  Suppose we fix the azimuthal quantum number $m$
and look for eigenfunctions of $\nabla^2(\nabla^2+2)$ with eigenvalue
$k$.  We can construct one by taking a linear combination
of two associated Legendre functions $P_{\lambda_+ m}$ and $P_{\lambda_- m}$,
where $\lambda_\pm$ are the two roots of
\be
(-\lambda_{\pm}(\lambda_\pm+1)+1)^2=k+1.
\ee
The left-hand side is the eigenvalue of $(\nabla^2+1)^2$; recall
that the eigenvalue associated with $P_{lm}$ is $-l(l+1)$, and compare
this equation to equation (\ref{eq:factorization}).

Just as in the case of the disk, there will be a discrete set of $k$'s
for which a linear combination of these two functions can satisfy
both boundary conditions.


\section{Pixelized Maps}

In this section we study the decomposition of polarization in finite
pixelized maps.  One possibility would be to search for eigenfunctions
of a discretized version of the bilaplacian operator.  On scales much
larger than the pixel scale, we would expect to recover modes that are
approximately the same as those found above.  The orthogonality of
pure $E$ and $B$ modes would not be expected to be perfect in the
discretized case, but on reasonably large scales it should be
close. The main drawback of this approach is that by construction 
it explicitly assumes that both $Q$ and $U$ are measured at each pixel,
so we would like to generalize the approach preserving its spirit and power.

We will adopt a different method in which a complete set
of E, B, and ambiguous modes can all be found at once by solving a
single eigenvalue problem.  With this approach, we can find a basis of
modes that approximate the pure $E$ and $B$ modes very well (except for
modes with frequencies close to the Nyquist frequency, where problems
may be expected to arise no matter what approach one adopts). 

We assume that we have a map of a finite portion
of the sky composed of $N$ pixels.  In each pixel we could have
measured both $Q$ and $U$; however, it is possible that in some or all
of them only one combination of the Stokes parameters was measured.
We will denote the vector of measured Stokes parameters $\P$, which
will have dimension less than or equal to $2N$. In terms of the $E$ and
$B$ modes of the full sky, the vector $\P$ is given by:
\beq \P=-\sum_\lm
(a_\elm \ \Y_\elm + a_\blm \ \Y_\blm). 
\eeq 

Ideally we want to find the pixelized analogues of the pure $E$,
pure $B$ and ambiguous modes.  A pure $E$ mode, which we will denote
$\e$, should satisfy $\Y_\blm \cdot \e=0$ for all $lm$. A pure $B$ mode
$\b$ satisfies $\Y_\elm \cdot \b=0$ for all $lm$. 
It is clear that one cannot find 
a solution to these sets of equations, 
\ie \ to find such an $\e$ or $\b$, since in general, we are
trying to satisfy more equations than we have components of the $\P$
vector. In practice the number of constraints we need to satisfy is
set by the angular resolution of the experiment, which determines the
maximum $l$ mode that has any appreciable power. Thus the
difficulty of finding pure $\e$ or pure $\b$ modes will increase as
the distance between pixels gets larger compared to the angular
resolution of the experiment. Moreover, we also expect that the
number of pure $E$ and pure $B$ modes will decrease as the fraction 
of pixels where only one of the Stokes parameters is measured
increases.
 
A pure $E$ mode should satisfy
\beq 
\sum_\lm C_{Bl} || \Y_\blm \cdot \e ||^2 = 0\ \ ,
\eeq
or equivalently
\beq
\e^t \B \e = 0 \ \  ; \ \ \B= \sum_\lm C_{Bl}  \Y_\blm \Y^{\dagger}_\blm,
\eeq
for any choice of power spectrum $C_{Bl}$.  An analogous statement
clearly holds for pure $B$ modes:
\beq
\b^t \E \b = 0 \ \  ; \ \ \E= \sum_\lm C_{El}  \Y_\elm \Y^{\dagger}_\elm 
\eeq 
for any $C_{El}$.  The matrices $\E$ and $\B$ give the contribution to
the power in each mode from the $E$ and $B$ components. In order to
find candidate $E$ and $B$ modes numerically, we must choose a particular
power spectrum; we will choose 
$C_{(E,B)l}/2\pi=(l-2)!/(l+2)! \times\ W_l^2$, where $W_l^2$ is
the window function that describes the beam smearing. We will motivate
this choice in the next subsection: in
practice we found it to work extremely well, making mixing between modes
extremely small and almost perfectly recovering the modes we obtained in
the previous section with the bilaplacian.

Our aim is to construct a basis of vectors that span all the space but
are ordered by their relative contributions from $E$ and $B$ modes. 
In principle, we would like to find the generalized eigenvectors of 
something like $\E \e= \lambda_E \B \e$.  A problem arises, however: 
we know that $\B$ has a null space (the space of
pure $E$ modes). So we regularize the problem by introducing a matrix
$\N=\sigma^2 \I$, with $\I$ the identity matrix and
$\sigma^2$ a very small constant. We then solve
\beq\label{eqe}
(\E+\N) \e= \lambda_E (\B+\N) \e.
\eeq
If we choose $\sigma^2$ small enough, the matrix $\E+\N$ is
essentially equal to to $E$ in the subspace of pure $E$ modes and is
proportional to the identity matrix in the subspace of pure $B$
modes. The converse holds for $\B+\N$. As a consequence, the
eigenvectors with large $\lambda_E$ 
will be very close to pure $E$ modes. Furthermore, with our choice of
power spectra, the eigenvectors will
automatically separate in scale with larger scale modes having
a larger eigenvalue.  

There is an equivalent equation for $B$ modes,  
\beq\label{eqb}
(\B+\N) \b= \lambda_B (\E+\N) \b,
\eeq
but any mode $\e$ satisfying equation (\ref{eqe})
also satisfies equation (\ref{eqb}) with $\lambda_B=1/\lambda_E$. 

We can derive simple and useful properties of the eigenvalues and
eigenvectors if we assume that at every pixel in the map we have
both $Q$ and $U$. We consider the simple transformation where 
we rotate the polarization at every pixel by $45^\circ$ (\ie \ $Q
\rightarrow -U$ and $U \rightarrow Q$). We denote this
transformation $\RR$. It is represented by a block diagonal
matrix  
\beqa
(\RR)_{ij}&=& \delta_{ij}
\left(
\begin{array}{cc}
0 & -1 \\
1 & 0 \\
\end{array}
\right),
\eeqa
where $i,j$ label pixels. The matrices $\E$ and $\B$ satisfy $\RR^t \E
\RR = \B$ and $\RR^t \B \RR = \E$. Moreover $\RR^t \RR = \I$. By
substitution into equation (\ref{eqe}), it is straightforward to
prove that the vector $\e^{\prime}=\RR\e$ also solves the eigenvalue
equation but with eigenvalue $1/\lambda_E$.  We conclude that if at
every pixel we have measured both $Q$ and $U$, modes that solve
equation (\ref{eqe}) come in pairs with eigenvalues $\lambda_E$ and
$1/\lambda_E$. One member of the pair is preferentially $E$ and the
other preferentially $B$.

In the next section we will present numerical examples to
gain intuition on how the eigenvalue problem works. 
First we will motivate our choice of $C_l$ spectra.

\subsection{Relation to bilaplacian formalism}

To find the relation between our eigenvalue and bilaplacian 
formalisms, we start by considering a vector satisfying the eigenvalue equation,
\beq\label{rel1}
(\E+\Nid) \e= \lambda_E (\B+\Nid) \e. 
\eeq
Since $\D_E^\dag\B=\D_B^\dag\E\b=0$, multiplying \eq{rel1}
by $\D_E^\dag$ and $\D_B^\dag$ yields two scalar equations,
\beqa\label{deigen}
\D_E^\dag \E \e &=& \sigma^2(\lambda_E-1)\D_E^\dag \e, \nonumber \\
\D_B^\dag \B \e &=& \sigma^2(\lambda_E^{-1}-1)\D_B^\dag \e. 
\eeqa
We now proceed to show that with our choice of spectra $C_{El}=
(l-2)!/(l+2)!$ the modes constructed using our bilaplacian formalism
solve equation (\ref{deigen}). We take 
\beq
\nabla^2(\nabla^2+2) \psi_E = \lambda \psi_E.
\eeq
and assume that $\psi_E$ satisfies both Dirichlet and Neumann boundary
conditions. We can use the completeness relation for spherical
harmonics and our choice of spectra to write
\beqa
\lambda^{-1}\nabla^2(\nabla^2+2) \psi_E(\th)&=&
\int d\th^\prime 
\sum_\lm 
\Yo_{\lm}(\th) \Yo_{\lm}^*(\th^\prime)  \psi_E(\th^\prime) \nonumber \\
&=&
\int d\th^\prime 
\sum_\lm C_{El}  {(l+2)!\over (l-2)!} 
\Yo_{\lm}(\th) \Yo_{\lm}^*(\th^\prime)  \psi_E(\th^\prime) 
\eeqa
We can use the fact that $ \D_E^{\dagger}  \Y_{E,\lm} =
[(l+2)!/(l-2)!]^{1/2} \Yo_\lm$ to get
\beq
\int d\th^\prime 
\sum_\lm C_{El}  
\D_E^\dagger \Y_{E,\lm}(\th) \Y_{E,\lm}^\dagger(\th^\prime)\  \D_E 
\psi_E(\th^\prime) =  
\lambda^{-1} \nabla^2(\nabla^2+2) \psi_E(\th)
\eeq
where we have integrated by parts using the boundary conditions
satisfied by $\psi_E$. Finally we can factorize the bilaplacian
operator $\D_E^\dag\cdot\D_E=\nabla^2(\nabla^2+2)$ and use our definitions 
$\e=\D_E \psi_E$ and the $\E$ matrix to get
\beq
\D_E^\dag \E \e = \lambda^{-1} \D_E^\dag \e.
\eeq
Thus if we identify $\lambda^{-1}= \sigma^2 (\lambda_E - 1)$,
$\D_E\psi_E$ satisfies the first of equations
(\ref{deigen}).  The second equation in (\ref{deigen}) is trivially
satisfied because being $\e$ a pure $E$ mode it follows that 
both $\B\e=0$ and $\D_B^\dag \e=0$. 

%

We have just shown that modes constructed using the bilaplacian formalism solve equation 
(\ref{deigen}) rather than (\ref{rel1}). This means that the vector $\e=\D_E\psi_E$ actually satisfies 
\beq\label{rel2}
(\E+\Nid)\e = \lambda (\B+\Nid) \e + \a,
\eeq
where $\a$ has to be an ambiguous mode (because it has to give zero when acted upon by both $\D_E^\dag$ and $\D_B^\dag$).  The easiest way to understand what is happening is to look at the structure of the $\E$ and $\B$ matrices in the basis of the eigenfunctions of the bilaplacian. If we call $\al$ one of the basis vector in the ambiguous space and contract equation (\ref{rel2}) with it we find,
\beq
\al^t\E\e=\al \cdot \a,
\eeq
where we have also used the fact that $\e$ was a pure $E$ mode.
Thus the reason why there is an extra ambiguous mode in eq. (\ref{rel2}) is that the $\E$ matrix can have non-zero elements mixing the pure $E$ and ambiguous subspaces. In other words our two formalisms are identical when restricted to the pure $E$ and $B$ subspaces but differ in the ambiguous subspace.

In practice we will find that the modes calculated by solving the
bilaplacian equation and the generalized eigenvalue problem are almost
identical. This can be understood by looking at \eq{rel2} and
realizing that in most cases we will be able to achieve very good
separation, {\it i.e.}, $\lambda \gg 1$.  This implies that one only
needs to add a very tiny amount of ambiguous modes to $\e$ in equation
(\ref{rel2}) to ``correct it'' and make $\a$ zero (because $\lambda$ is
so large). This is especially so because under most circumstances the
matrix elements of both $\E$ and $\B$ in the subspace of ambiguous
modes are comparable.

\section{Worked examples II}

We begin by revisiting the cap example we solved in the continuous
case. We start by assuming that every pixel has both $Q$ and $U$.  We
consider a fiducial experiment with a $0.2$ degrees FWHM for the beam
angular resolution. The patch observed has a radius of $3.8$ degrees
and contains 351 pixels (the spacing between pixels in both the radial
and the tangential directions was set to 0.2 degrees as well). 


\begin{figure}[ht]
\centerline{\epsfxsize=13cm\epsffile{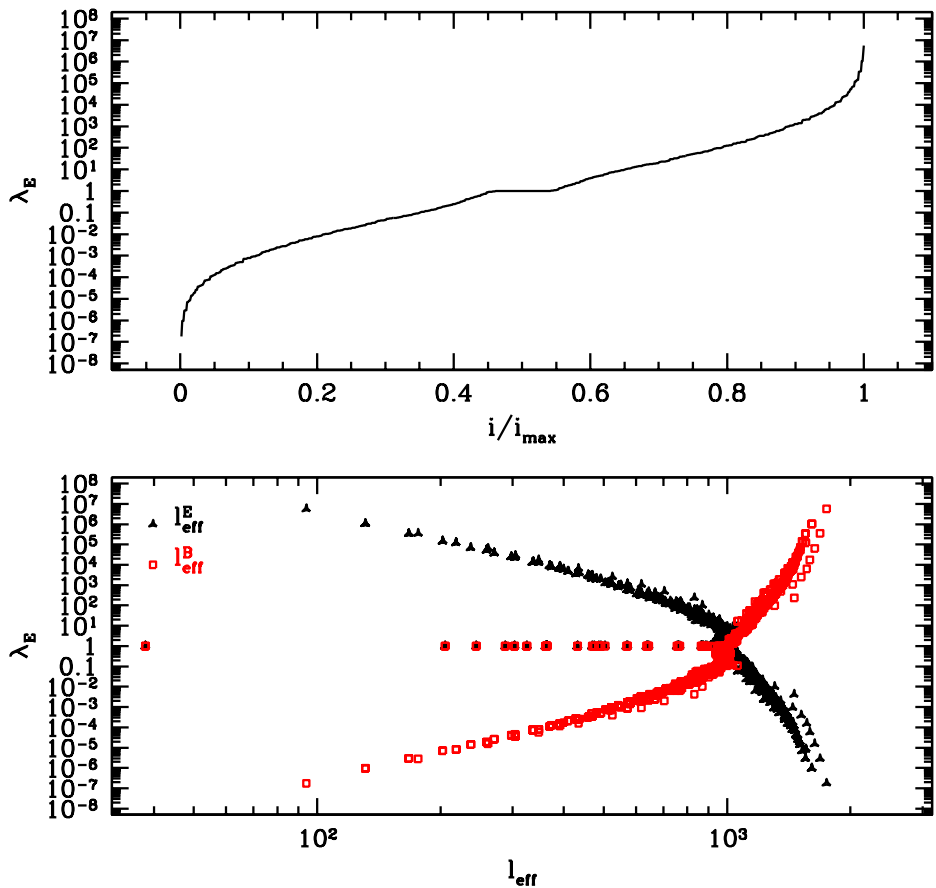}}
\caption{$E/B$ eigenvalues for a cap. In the top panel we show
$\lambda_E$ as a function of mode number. On the bottom we show the
eigenvalues as a function of both $\leff^E$ and $\leff^B$ as defined
in equation (\ref{eq:leff}). We took $\sigma^2=4\times 10^{-6} $, a factor $10^{-5}$ smaller than the zero lag correlation function.}
\label{lb1}
\end{figure}

Figure \ref{lb1} shows the eigenvalues we obtained. As expected, the
eigenvectors come in pairs with eigenvalues $\lambda_E$ and
$1/\lambda_E$. The eigenvectors with very small eigenvalues correspond
to pure $B$ modes and those with very large ones to pure $E$
modes. The particular values of the eigenvalues should not be given
much importance as they depend on value of the regularizing constant
$\sigma^2$. What is important is that the large eigenvalues show the
good degree of separation that we have achieved.

There is also a
concentration of modes at $\lambda=1$.  These modes have two
origins. First, modes on small scales, where our small $\sigma^2 {\bf
I}$ regularization dominates over the $\E$ and $\B$ matrices, will
have $\lambda_E=1$. Second, as we discussed in the previous sections,
there are larger-scale ambiguous modes that receive contributions 
from both $E$ and $B$.  Our method is unable to
separate between both types because they have the same eigenvalues.

\begin{figure}[ht]
\centerline{\epsfxsize=13cm\epsffile{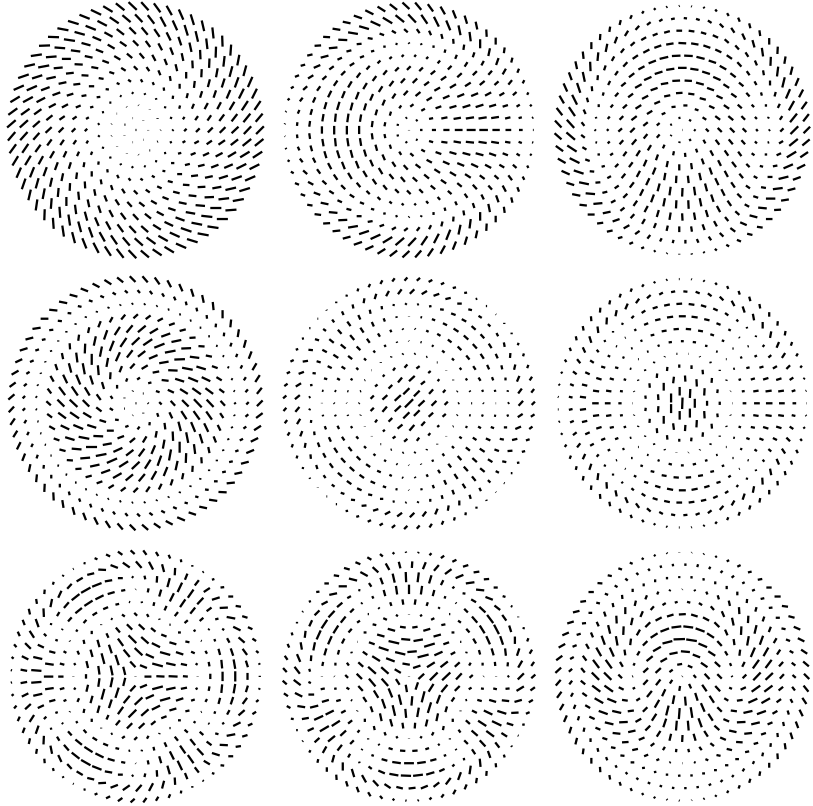}}
\caption{Examples of modes for a cap. We show the nine modes with
lowest $\lambda_E$.}
\label{modes1}
\end{figure}

In Figure \ref{modes1} we show the first nine eigenvectors, corresponding
to the lowest nine eigenvalues. One immediately recognizes in this set the
pure $B$ modes discussed is section \ref{disk}. The first eigenvector
corresponds to the lowest-order $m=0$ mode. The next two are the lowest
$m=1$ modes, which differ only by a rotation.
Then come the second $m=0$ mode, then the lowest $m=2$ modes, then the
lowest $m=3$ modes, and finally the second $m=1$ mode. 
The best nine $E$ modes,
corresponding to the largest nine eigenvalues, are simply equal to the
ones plotted in figure \ref{modes1} but with each polarization ``vector''  
rotated by $45^\circ$. 

\begin{figure}[ht]
\centerline{\epsfxsize=13cm\epsffile{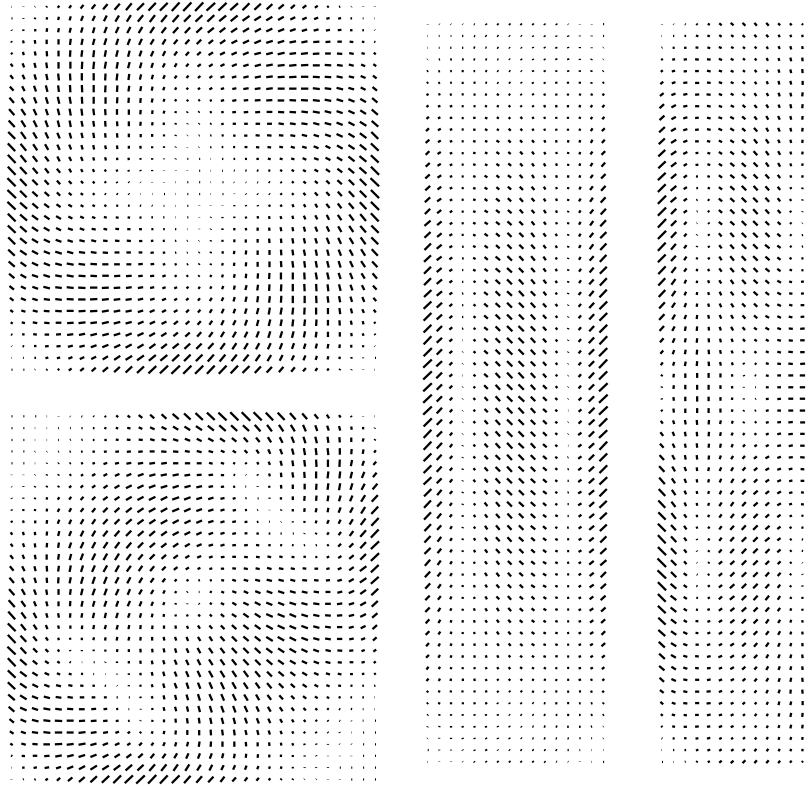}}
\caption{On the left (right) we show the best two $B$ type eigenvectors for a
$32\times 32$ ($16\times 64$) pixel patch.}
\label{sqpathc}
\end{figure}

Our method for finding modes can be used for any shape of sky
patch. In figure \ref{sqpathc} we show the first two modes of a square
patch $32\times 32$ pixels on a side.  Comparing with figure
\ref{modes1}, it is clear that they are essentially the same modes as
the two first modes for the cap. We also show the first two modes in a
patch $16\times 64$ pixels on a side. We have also checked that these
modes for the rectangle can be derived from the bilapacian formalism.

To understand where the ordering of modes in figure \ref{modes1} is
coming from, {\it i.e.}, why the modes appear in that order in the
figure, we will introduce window functions for each mode. We
define
\beqa
W^{E}_{l}&=&\e^t {\partial \E \over \partial p_{l}}\e \\
W^{B}_{l}&=&\e^t {\partial \B \over \partial p_{l}}\e, 
\eeqa
where we have introduced $p_{l}=l(l+1)C_l/2\pi$. 
Using the window functions we can define an effective $l$ for each
mode, the average $l$ calculated using the window function as a
weight. Specifically, we can define quantities
\beq
l^{(E,B)}_{\mbox{\scriptsize eff}}={\sum_l \ l \ W_l^{(E,B)}\over\sum_l 
\ W_l^{(E,B)}}
\label{eq:leff}
\eeq
that give
the average $l$ for the $E$ and $B$ contribution to a given mode.

\begin{figure}
\centerline{\epsfxsize=13cm\epsffile{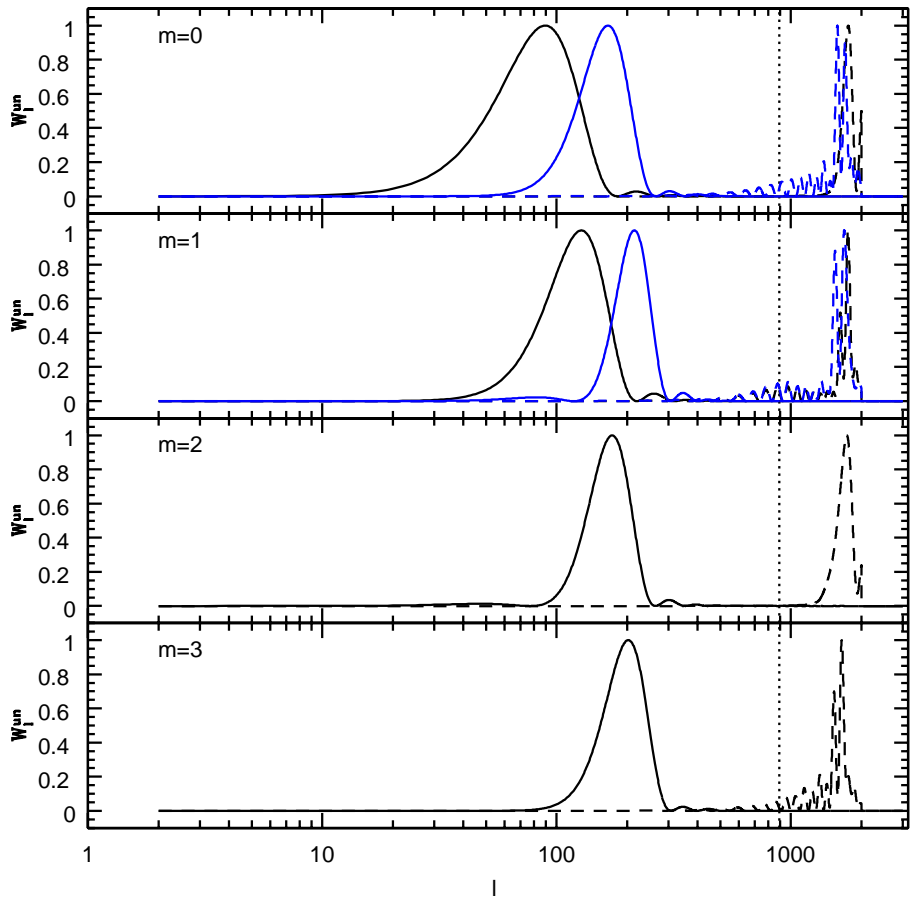}}
\caption{Window functions for the modes shown in figure
\ref{modes1}. Solid lines show $W_l^B$ and dashed lines show
$W_l^E$, normalized to unit peak height. }
\label{win1}
\end{figure}

Figure \ref{win1} shows the window functions for the eigenvectors
that were plotted in figure \ref{modes1}.  
Note that the window functions are well localized in $l$ and each has a clear
peak. Moreover, the modes in figure \ref{modes1} are ordered in
increasing order of $\leff^B$. Both of these are a consequence of
our choice of power spectra.

\section{Aliasing}

The windows $W_l^E$ in figure \ref{win1} can be used to determine
where the leakage between $E$ and $B$ is coming from.  The dotted
line in all the panels gives an estimate of the Nyquist frequency in
the map. The conclusion is clear: the contaminating power is aliased
power. For power that is aliased one cannot
distinguish $E$ from $B$. The remedy for this is to increase the
sampling in the map so as to further suppress the aliased power.  

The bottom panel of figure \ref{lb1} shows the eigenvalues we obtained
for the cap but as a function of $\leff^E$ and $\leff^B$. We see
that the modes with large values of $\lambda_E$ have a low $\leff^E$
and a large $\leff^B$, indicating that most of the contamination is
coming from aliasing. The opposite is true for modes with low
$\lambda_E$. We also see that some of the modes with $\lambda_E=1$
receive contributions from large scales, an indication that these
are truly ambiguous modes.

We can understand our results intuitively
by considering a simple toy model, closely following the treatment in \cite{ted}. 
We work in the small-angle limit and compute 
the Fourier components of the observed polarization
field assuming they were observed over a square patch of size
$L$. Using equation (\ref{fourier}) we
obtain
\beq
\tilde \P(\k) 
= \int {d^2\q \over (2\pi)^2} W(\k-\q) \left[E(\q ) 
\left(\matrix{\cos (2\phi)  \cr \sin (2\phi)}
\right) + B(\q) \left(\matrix{- \sin (2\phi)  \cr \cos (2\phi)}
\right)\right],
\eeq
where we have defined the window function 
\beq W(\k)=\left({\Delta \theta
\over L}\right)^2\  {\sin k_x L /2 \over \sin k_x \Delta\theta /2 }\ 
{\sin k_y L /2 \over \sin k_y \Delta\theta /2 },
\eeq 
with $\Delta\theta$ the separation between pixels. The Nyquist
wavenumber is $k_{\mbox{\scriptsize Nyq}} = \pi/\Delta \theta$. 

The naive way to recover the $E$ and $B$ components would be to combine
the Fourier coefficients  $\tilde \P$ as one would do if the patch were
infinite. For $\tilde B$ for example, we would compute
\beq
\tilde B(\k)
=-\sin 2 \phi \ \tilde Q(\k) + \cos 2 \phi \ \tilde U(\k)  
\eeq 
and then estimate the power spectrum by taking the square of these
variables.

In terms of the real $E$ and $B$, our $B$-estimate can be written as
\beq
\tilde B(\k) = \int {d^2\q \over (2\pi)^2} W(\k-\q) [ \sin 2
\alpha \ \ E(\q) + \cos 2 \alpha \ \ B(\q)], 
\eeq 
where $\cos\alpha=\k \cdot \q /k q$. This estimate has contributions from both
$E$ and $B$. Only when $W$ is a delta-function such that $\alpha=0$ do we
avoid mixing. The $E$ contributions arise because of two effects:
the finite size of the sky patch and the pixelization (causing aliasing).  
The effect of the finite patch
size manifests itself as a finite width of the peaks
of the window function, while the effect of aliasing is that $W$
has several peaks.

We have shown how to construct modes that avoid contamination due to
the finite patch size. These modes are not Fourier modes.
In what follows we want to show that even in the limit $L \rightarrow
\infty$, there is still mixing due to pixelization. If we take this
limit the window function becomes a sum of delta functions centered at
$\q=2(m,n)k_{\mbox{\scriptsize Nyq}}$, where $m$ and $n$ are integers. 

To consider a concrete example, we calculate the ratio of power in $E$
and $B$ in a mode with wave vector $\k_0$ produced by an initial field
that only had $E$ modes with a power spectrum $C_l$ 
and assuming an infinite but pixelized sky map.
All the wave vectors $\k_{ij}=\k_0+2(i,j)k_{\mbox{\scriptsize Nyq}}$ will
contribute to this mode. We get
\beq
{\expec{\tilde B(\k_0)}\over\expec{\tilde E(\k_0)}}
={\sum_{ij} \sin^2 2(\phi_{ij} - \phi_0) C_{l_{ij}}/C_{l_0} \over 
\sum_{ij} \cos^2 2(\phi_{ij}-\phi_0) C_{l_{ij}}/C_{l_0}},
\label{ratio}
\eeq
where $l_{ij}\equiv |{\bf k}_{ij}|$ and $l_0\equiv |{\bf k}_0|$.
Equation (\ref{ratio}) shows that all the aliased modes contribute to
$B$ contamination because in general these modes do not have
$2(\phi_{l_{ij}} - \phi_{l_0})=m\pi$.  

It is important to note that the aliased power is suppressed by the
beam. As $i$ and $j$ become larger, the magnitude of the power on
those scales decreases because $C_l$ is proportional to $W_l^2$, the
beam window function. For example, if we consider a mode with
wavevector along the positive $x$ axis, 
the aliased mode with the smallest possible beam suppression has a
power suppressed by a factor $\exp[-2 k_{\mbox{\scriptsize
Nyq}}(k_{\mbox{\scriptsize Nyq}}-| {\bf k}_0|)/\sigma_b^2]$, where
$\sigma_b$ is the Gaussian width of the beam ($\sigma_b$ is related to
the full width half max of the beam (FWHM) by $\sigma_b=\sqrt{8
\ln(2)}/{\rm FWHM}$). For fixed $\k_0$, the suppression can be made as
large as one wants by increasing $k_{\mbox{\scriptsize Nyq}}$, that is
by increasing the sampling of the map.  If we want the beam to produce
a suppression factor $S$ we need to choose $2 k_{\mbox{\scriptsize
Nyq}}^2(1-| {\bf k}_0|/k_{\mbox{\scriptsize Nyq}})/\sigma_b^2]=\ln(S)$,
or equivalently ${\rm FWHM}/\Delta\theta \approx 0.5 \sqrt{\ln(S)/(1-|
{\bf k}_0|/k_{\mbox{\scriptsize Nyq}})}$.

\begin{figure}
\centerline{\epsfxsize=13cm\epsffile{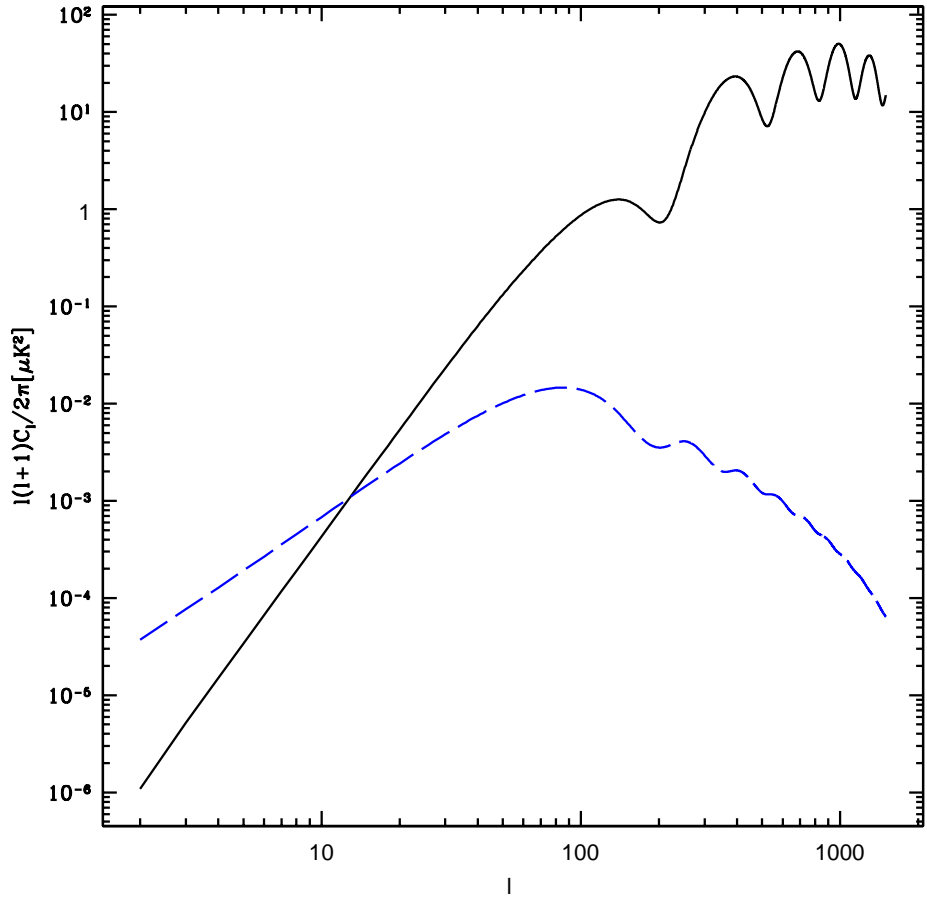}}
\caption{Polarization power spectra in a $\Lambda$CDM model. The
solid curve is for $E$ produced by density perturbations and the
dashed curve for the $B$ component produced by tensor modes. 
The anisotropies were COBE normalized and it was assumed that the tensor
component was $10\%$ of the anisotropies on these scales. }
\label{cls}
\end{figure}

A point worth noting about aliasing is that the power spectrum of the
polarization is a rapidly growing function of $l$ and that the $E$
power spectrum is expected to be much larger than the $B$ one. Figure
\ref{cls} shows the power spectrum for $E$ and $B$ type polarization
from gravity waves in a $\Lambda$CDM model. The temperature spectra
were COBE normalized and the tensor component was assumed to be 10\%
of the temperature anisotropies on COBE scales. The sharp increase in
power between $E$ and $B$ partially compensates the smearing by the
beam.  To give a rough feeling of what sampling is needed to avoid
aliasing we could assume that we want the aliased power to be a factor
of 100 smaller than the power we want to measure. For a temperature
map that would correspond to a suppression factor $S \sim 10^{2}$
while for polarization we would need $S \sim 10^{5}$ which means that
the ratio ${\rm FWHM} /\Delta \theta$ has to be a factor $\sqrt{2.5}$
larger for polarization than for temperature, or equivalently that one
needs a factor of $2.5$ more pixels to obtain the same level of
contamination.  We conclude that one has to be particularly careful
about aliasing when dealing with polarization maps if one wants to
obtain a clean separation between $E$ and $B$.

\begin{figure}
\centerline{\epsfxsize=13cm\epsffile{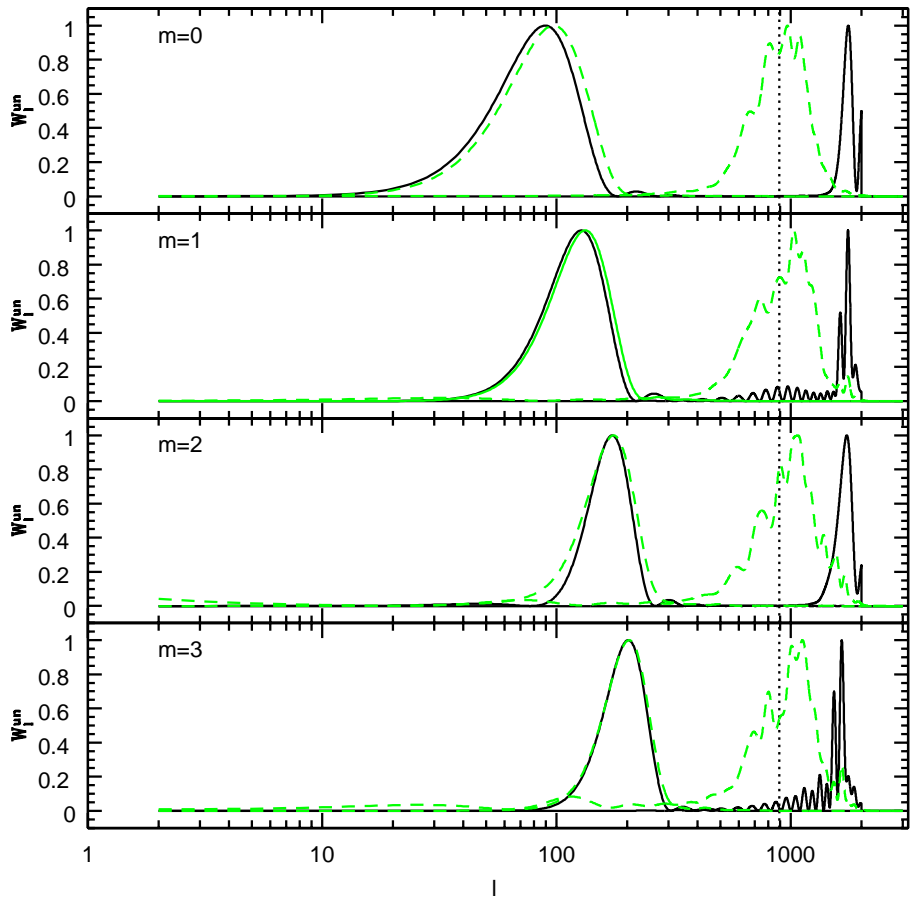}}
\caption{Effect of measuring only one Stokes parameter on 20\% of the
pixels chosen at random. The panels show the window functions for the
first modes with $m=0,1,2,3$ when both Stokes parameters are measured
(solid lines) and when 20\% are missing (dashed lines).}
\label{randomwins}
\end{figure}

The effect of aliasing can be decreased by increasing the sampling of
the map. It should be noted, however, that the presence of holes, bad
pixels or pixels with only one measured Stokes parameter
in the map will have a similar effect. We illustrate  this by considering a toy
example. We artificially increase the noise variance (the diagonal
elements of $\N$) for a fraction of the pixels chosen at random. Figure
\ref{randomwins} shows the window functions for the first nine modes
in an example where $20 \%$ of the measured Stokes parameter were
assumed to have the large noise. For comparison, we also show the
original window functions. On large scales, the modes look
essentially the same as the ones plotted in figure \ref{modes1}.  The
effect of the missing pixels is very noticeable in the $E$ window
function, the one that quantifies the leakage. As might have been
expected, the level of contamination coming from modes of frequency
around the Nyquist frequency is greatly increased. 

\section{Discussion}

We have developed a formalism for measuring the $E$ and $B$ components of 
polarized CMB maps or weak lensing maps given the real-world complications of 
finite sky coverage and pixelization. 

We have shown that by expanding a map in a particular basis, obtained
by differentiating bilaplacian eigenfunctions, it can be decomposed as
a sum of three orthogonal components that we term pure $E$, pure $B$
and ambiguous.  The pure $E$-component is orthogonal to all $B$-modes
and are therefore guaranteed to be caused by an $E$ signal (on the
uncut sky), and conversely for the pure $B$-component.  The ambiguous
component is the derivative of a biharmonic function, and the original
map contains no information about whether it is due to $E$- or
$B$-signal in the uncut sky.  We also derived a discrete analogue of
these results, applicable to pixelized sky maps.  Our results are
useful both for providing intuition for survey design and for
analyzing data sets in practice.

\subsection{Implications for survey design}


To maximize our ability to separate $E$ and $B$, 
we clearly want to minimize the fraction of modes that are ambiguous. 
We found that the ambiguous modes are specified along the boundary
of the map rather than in the two-dimensional interior. This means that
the number of pure and ambiguous modes probing a characteristic 
angular scale $\theta$ scales as the map area over $\theta^2$ 
and as the map boundary length over $\theta$, respectively.
It is therefore best to minimize the ratio of circumference to area, i.e., to 
make the patch as round as possible.

Almost all pure modes  (all except the ones with $m=0$ for the
spherical cap example) are a combination of both $Q$ and $U$ Stokes
parameters, so to achieve unambiguous $E/B$ separation, one needs to measure
both, with comparable sensitivity throughout the map.

With pixelized maps, we found that aliasing of small-scale power was a
serious problem. Although it can in principle be eliminated by heavily
oversampling the map, the required oversampling is greater than for
the unpolarized case, both because derivatives are involved and
because CMB polarization is expected to have an extremely blue power
spectrum. This has important implications for, e.g., the Planck
satellite, where bandwidth constraints on the telemetry have been
mentioned as reasons to reduce the oversampling.  It is crucial to
bear in mind that the usual Nyquist rule-of-thumb that applies to
unpolarized maps may be insufficient for realizing the full scientific
potential of Planck's CMB polarization measurements because one needs
roughly a factor of 2 to 3 more pixels in a polarization map to
achieve the same level of contamination by aliased power.

\subsection{Implications for data analysis}

In \cite{maxangelica}, it was shown how a quadratic estimator method
could produce uncorrelated measurements of the $E$ and $B$ power
spectra from real-world data sets with arbitrary sky coverage,
pixelization and noise properties, and this method has been applied to
both the POLAR \cite{Keating01} and PIQUE \cite{angelica} data.  The
one annoying problem with this method was that it gave $E/B$ leakage.
Our present results allow us to understand and eliminate this problem.

We now know that leakage is caused by the ambiguous modes.  The
above-mentioned scaling tells us that the fraction of modes probing a
given angular scale $l\sim \theta^{-1}$ that are ambiguous scales as
$l^{-1}$, in good agreement with the asymptotic behavior empirically
found in \cite{maxangelica}.  Although \cite{maxangelica} presented a
technique for removing most of the leakage, we now know how to remove
it completely: by eliminating the ambiguous modes.

In practice, the way to do this is to compute two projection matrices
$\PP_E$ and $\PP_B$ that project onto the subspaces given by the
eigenvectors $\e$ of \eq{eqe} with $\lambda_E>\lambda_*$ and
$\lambda_E<1/\lambda_*$, respectively, for some large eigenvalue
cutoff $\lambda_*$, say $\lambda_*=100$.  The three maps $\PP_E\P$,
$\PP_B\P$ and $[\I-\PP_E-\PP_B]P$ will then be approximately the pure
$E$, pure $B$ and ambiguous components of the original map $\P$, which
can be directly used for visual inspection, cross-correlation with
other maps and systematic error tests.  To measure the $E$ and $B$
power spectra, one compresses the original data vector $\P$ into two
shorter ones $\P_E$ and $\P_B$ by expanding it into the
above-mentioned pure $E$ and pure $B$ eigenvectors,
respectively. Since this is a mere matrix multiplication, the
corresponding noise and signal covariance matrices (which the
quadratic estimation method takes as input) are trivially computed as
well.  These two data vectors will each have less than half the length
of $\P$. Since the time required by the quadratic estimator method
scales as $n^3$, the final $E$ and $B$ power spectrum calculations are
therefore about an order of magnitude faster than in the original
\cite{maxangelica} approach.

It should be noted that the ambiguous modes are not useless in all
circumstances.  If it has been established that $E$ dominates over $B$
(as is expected theoretically) by observing the pure modes, then it is
safe to assume that most of the power in the ambiguous modes is $E$
power as well.  In this case, the ambiguous modes can be used to
reduce the errors on estimates of the $E$ power spectrum. This could
be particularly useful when attempting to constrain reionization with
$E$-power on the very largest angular scales attainable with a
galaxy-cut all-sky map, where a substantial fraction of the modes will
be ambiguous.

\vskip 1cm
{\bf Acknowledgments:} 
The authors thank Ue-Li Pen for asking questions that stimulated this work.
Supported was provided by
NSF grants AST-0071213, AST-0134999, AST-0098048, AST-0098606 and PHY-0116590,
NASA grants NAG5-9194 \& NAG5-11099,
and two Fellowships from the David and Lucile Packard Foundation.
MT and EFB are Cottrell Scholars of the Research Corporation.


\begin{references}
%
%

\bibitem{Keating01}
Keating, B. et al., ApJL, 560, L1 (2001)

\bibitem{staggs}
S.T. Staggs, J.O. Gundersen, \& S.E. Church, in {\it Microwave Foregrounds},
edited by A. de Oliveira Costa and M. Tegmark (ASP Conference Series,
vol. 181, San Francisco), p. 299.

\bibitem{hedman}
M.M. Hedman, D. Barkats, J.O. Gundersen, S.T. Staggs, \& B. Winstein,
\apj Lett. {\bf 548}, L111 (2001).

\bibitem{peterson}
J.B. Peterson, J.E. Carlstrom, E.S. Cheng, M. Kamionkowski, A.E. Lange,
M. Seiffert, D.N. Spergel, \& A. Stebbins,
astro-ph/9907276 (1999).

\bibitem{angelica}
A. de Oliveira-Costa, M. Tegmark, M. Zaldarriaga, D. Barkats, 
J.O. Gundersen, M.M. Hedman, S.T. Staggs, \& B. Winstein,
astro-ph/0204021 (2002).


\bibitem{zalreio} M. Zaldarriaga, 
\prd {\bf 55}, 1822 (1997).

\bibitem{peebles}
P.J.E. Peebles, S. Seager, \& W. Hu, \apj Lett. {\bf 539}, L1 (2000).

\bibitem{landau}
S.J.Landau, D.D. Harari, \& M. Zaldarriaga \prd {\bf 63}, 3505 (2001).

\bibitem{sperzal} D.N. Spergel
\& M. Zaldarriaga, \prl {\bf 79}, 2180 (1997).

\bibitem{2.kks} M. Kamionkowski, A. Kosowsky \& A. Stebbins,
\prd, {\bf 55}, 7368 (1997).

\bibitem{3.spinlong} 
M. Zaldarriaga \& U. Seljak, \prd {\bf 55}, 1830 (1997).

\bibitem{spinlett} U. Seljak
\& M. Zaldarriaga, \prl  {\bf 78}, 2054 (1997).

\bibitem{kkslett} M. Kamionkowski, A. Kosowsky \& A. Stebbins,
\prl, {\bf 78}, 2058 (1997).

\bibitem{kinney} 
W.H. Kinney, \prd  {\bf 58}, 123506 (1998).

\bibitem{pollens}
M. Zaldarriaga \& U. Seljak, \prd {\bf 58}, 3003 (1998).

\bibitem{jacek}
J. Guzik, U. Seljak, \& M. Zaldarriaga, \prd, {\bf 62}, 3517 (2000).

\bibitem{karim} K. Benabed, F. Bernardeau, \& L. van Waerbeke, 
\prd, {\bf 63}, 3501 (2001).

\bibitem{wayne}  W. Hu \& T. Okamoto, 
submitted to \apj, astro-ph/111606 (2001).

\bibitem{JKW}
A. Jaffe, M. Kamionkowski, and L. Wang,
Phys.Rev. D \textbf{61}, 083501 (2000).

\bibitem{maxangelica}
M. Tegmark and A. de Oliveira Costa, 
Phys. Rev. D, \textbf{64}, 063001 (2001).


\bibitem{ted}
E.F. Bunn, \prd, {\bf 65}, 043003 (2002).

\bibitem{lewis}
A. Lewis, A. Challinor, and N. Turok, astro-ph/0106536 (2001).


\bibitem{zal}
M. Zaldarriaga, 
\prd {\bf 64}, 103001 (2001).


\bibitem{huwhite}
W. Hu \& M. White, New Astronomy \textbf{2}, 323 (1997).


\bibitem{zalpol} 
M. Zaldarriaga, Astrophys. J. \textbf{503}, 1 (1998).

\bibitem{kaiserlens}
N. Kaiser, \apj \textbf{498}, 26 (1998).


\bibitem{huwhitelens}
W. Hu and M. White, 
\apj \textbf{554}, 67 (2001).

\bibitem{Crittenden02}
R. Crittenden,  P. Natarajan, U. L. Pen, and T. Theuns, \apj \textbf{568} 20 (2002)


\end{references}
\end{document}